\documentclass[preprint,12pt]{elsarticle}

\usepackage{longtable}
\usepackage{graphicx} 
\usepackage[linesnumbered,ruled,vlined]{algorithm2e}
\newcounter{magicrownumbers}
\newcommand\rownumber{\stepcounter{magicrownumbers}\arabic{magicrownumbers}}
\usepackage{xltabular}
\usepackage{supertabular}
\usepackage{tabularx}
\usepackage{enumitem}
\usepackage{amssymb}
\usepackage{amsmath}

\journal{Journal of Systems and Software}

\begin{document}

\begin{frontmatter}



\title{Beyond Pass/Fail: The Story of Learning-Based Testing}


\author{Sheikh Md. Mushfiqur Rahman}
\author{Nasir U. Eisty}

\affiliation{organization={University of Tennessee},
            addressline={Knoxville},
            state={TN},
            postcode={37996},
            country={USA}}

\begin{abstract}
Learning-Based Testing (LBT) merges learning and testing processes to achieve both testing and behavioral adequacy. LBT utilizes active learning to infer the model of the System Under Test (SUT), enabling scalability for large and complex programs by requiring only a minimal set of initial test cases. The core principle of LBT is that the SUT's behavior can be thoroughly inferred by progressively generating test cases and subjecting the SUT to testing, thereby ensuring comprehensive testing. Despite being in its early stages, LBT has a solid foundation of theoretical research demonstrating its efficacy in testing both procedural and reactive programs. This paper provides a systematic literature review of various LBT implementations across different program types and evaluates the current state of research in this field. We explore diverse theoretical frameworks, existing tools, and libraries within the LBT domain to illustrate the concept's evolution and current research status. Additionally, we examine case studies involving the application of LBT tools in industrial settings, highlighting their potential and effectiveness in commercial software testing. This systematic literature review aims to offer researchers a comprehensive perspective on the inception and development of LBT, presenting it as a promising technique in software testing. By unveiling LBT's underutilized potential, this paper seeks to significantly benefit the practitioners and research community.
\end{abstract}



\begin{keyword}
Learning Based Testing; Software Testing; Software Engineering

\end{keyword}

\end{frontmatter}



\section{Introduction}
Weyuker first introduced the concept of LBT in their Ph.D. research~\cite{weyuker1983assessing}. 
They portrayed testing as an inference process wherein testers try to discern software attributes by examining its response to distinct inputs. 
This approach operates under the premise that testing and model inference of a program are closely intertwined~\cite{walkinshaw2010increasing}. These inferred models reflect the testing coverage, allowing the test generator to seek new test cases that challenge the predictions made by the inferred models~\cite{rahman2025learningbasedtestingdeeplearning, 11154529, walkinshaw2010increasing}.

The LBT approach uses a concise test set for a System Under Test (SUT), labeled as $P$. 
It repeatedly produces a program, \( p' \) that's part of \( P \), conforming with the test set. 
This approach looks for a unique input differentiating \( P \) from \( p' \).  When such contradicting
test cases are found, they are added to the test set, and a new  \( p' \) program is inferred using the accumulated test set. 
This continues until only \( P \) can be derived from the generated examples \cite{bergadano1996testing}.
The essence of this technique lies in selecting test cases with a high likelihood of detecting faults by focusing on input/output relations that contradict the inferred model~\cite{papadopoulos2015black, walkinshaw2017uncertainty}.

LBT is essential because it significantly enhances the efficiency of testing large and complex software systems~\cite{walkinshaw2010increasing, romanik1996using}. Traditional testing methods often require extensive manual effort and large test suites, making them time-consuming and resource-intensive. LBT, on the other hand, achieves thorough testing with a minimal set of initial test cases through its active learning capabilities. This approach not only streamlines the testing process but also ensures that the SUT is rigorously evaluated.

The scalability of LBT is another critical factor that underscores its importance~\cite{walkinshaw2010increasing, quddus2022structural}. As modern software systems grow in complexity, the ability to infer the model of the SUT through active learning becomes invaluable. This scalability ensures that LBT can be effectively applied to a wide range of software, from small applications to large-scale enterprise systems. Furthermore, LBT's integration of learning and testing guarantees behavioral adequacy, meaning the SUT’s behavior is comprehensively understood and tested. This dual focus helps identify hidden defects that might not be captured by conventional testing methods.

Adaptability is also a key strength of LBT~\cite{bergadano1996testing, aichernig2020passive}. It can be tailored to test various types of software, including both procedural and reactive programs, making it a versatile approach suitable for diverse applications in software development. Additionally, LBT reduces the manual effort required for testing by automating much of the process. This automation allows developers and testers to concentrate on other critical tasks in the development lifecycle, thereby increasing overall productivity and efficiency.

In this paper, we conducted a comprehensive systematic literature review to reflect on various LBT methodologies, evaluating their efficacy both in controlled experiments and real-world scenarios. 
This review is crafted to provide researchers with an overarching perspective on LBT's evolution and its promising potential.
This review of LBT is essential for consolidating existing knowledge, identifying research gaps, and evaluating the effectiveness of LBT methodologies. By bringing together diverse research findings and practical implementations, this review provides a comprehensive overview of the current state of LBT, highlighting best practices and successful case studies. This paper not only raises awareness about the potential and advantages of LBT but also promotes its adoption in the software industry. Additionally, this paper serves as a state-of-the-art resource for future research, guiding efforts toward areas that need further exploration and fostering innovation in software testing.


\section{Background}

Algorithm~\ref{alg:generic_LBT} demonstrates a basic LBT algorithm. 
The algorithm begins with an initial collection of inputs $T_i$ alongside the SUT $P$ and a set of behavioral requirement specifications of the SUT $S$. $T_i$ might be empty, or it could be a preexisting test input that we aim to enhance. Depending on the specification set $S$, $OracleGen(input,S)$ function decides the expected output for any input in $T_i$. This input-output pair is added to the $TrainSet$ and is used to infer the model $M$. 

In the test generation loop, the first step is to infer a predictive input/output model, referred to as $M$, for the program using the function $inferModel(TrainSet)$.  The nature of this model can vary depending on the specific attributes of $P$. Then, $NewInputs$ are generated depending on the specification set $S$ to target specific attributes of $P$ or randomly. Then the $NewInputs$ are executed using the $M$, and the output of $M$ for the $NewInputs$ is used to select a set of test inputs using the function \textit{selection(M, S, Executions)}.  $OracleGen$ function generates the expected output for the $SelectedInputs$, and the input-output pairs are added to the final test set $T_f$. This selection process can be implemented in many ways: for example, inputs that are counterexamples could be selected, which means that for this input, the model's output is unexpected. Or, even if the model's output is expected, the model's confidence or surprise adequacy can be used to select the input. After selecting the inputs, the created new tests are added in the $TrainSet$ for inferring the model in the next loop.

The process of model inference and test generation continues until the termination criterion $terminate(T_f, M)$ is met. This criterion can vary; for instance, it may aim to verify the equivalence between the inferred model $M$ and the SUT $P$, returning true if the model and the SUT are deemed sufficiently similar in some predefined way. Alternatively, the loop could stop after a predetermined number of iterations once the final test set $T_f$ reaches a certain size or if there is no change in $M$ after a specified number of iterations. Finally, the algorithm returns the final test set $T_f$ to test on the SUT.

\begin{algorithm}
    \SetAlgoLined
    \SetKwInOut{Input}{\textit{Input}}
    \SetKwInOut{Uses}{\textit{Uses}}
    \SetKwInOut{Result}{\textit{Result}}
    
    \Input{SUT \textit{P}, Initial Test Input Set $T_i$, Specification $S$}
    \Uses{\textit{terminate}, \textit{OracleGen}, \textit{selection},
    \textit{execute},
    \textit{inferModel}}
    \Result{FInal Test Set $T_f$}
    
    \BlankLine
    Inferred Model  \textit{M} $\leftarrow \emptyset$\;
    \textit{TrainSet} $\leftarrow \emptyset$\;
    $T_f$ $\leftarrow \emptyset$\;
    \ForEach{\textit{input} $\in$ $T_i$}{
        \textit{TrainSet} $\leftarrow$ \textit{TrainSet} $\cup$ \textit{OracleGen(input,S)}\;
    }
    \While{(\textit{¬ terminate($T_f$, M, P)})}{
        \textit{M} $\leftarrow$ \textit{inferModel(TrainSet)}\;
        \textit{NewInputs} $\leftarrow$ \textit{testGenerator(S | random)}\;
        \textit{Executions} $\leftarrow$ \textit{execute(M, NewInputs)}\;
        \textit{SelectedInputs} $\leftarrow$
     \textit{selection(M, S, Executions)}\;
        \textit{TrainSet} $\leftarrow$ \textit{TrainSet} $\cup$ \textit{OracleGen(NewInputs,S)}\;
         $T_f$ $\leftarrow$ $T_f$ $\cup$ \textit{OracleGen(SelectedInputs,S)}\;
    }
    
    \Return $T_f$\;
    
    \caption{\textit{A general LBT algorithm}}
    \label{alg:generic_LBT}
\end{algorithm}

As we can see from  algorithm~\ref{alg:generic_LBT}, LBT addresses two core challenges of testing highlighted by Weyuker~\cite{weyuker1983assessing}: 

\begin{itemize}
    \item By pinpointing the counterexamples of the approximate model, LBT ensures the testing process picks only impactful test cases.
    \item By gauging the sufficiency of the inferred model, LBT decides the termination point.
\end{itemize}

LBT's incremental learning of the SUT sets it apart, enabling a scalable and efficient process~\cite{feng2013case}. The prime utility of the technique is its iterative construction of the SUT's approximate model, rendering it scalable from straightforward computational models to larger problems~\cite{meinke2018learning}. LBT's strength lies in its fluidity and adaptability, underscored by proactive learning and real-time testing. 
Such flexibility means it remains steadfast even amid unforeseen changes, like software refactoring. 
This resilience makes LBT perfectly aligned with modern agile development techniques, including continuous integration~\cite{meinke2018learning}. 

Although LBT is fundamentally a black-box testing technique, it addresses issues prevalent in source code-driven tests, notably that code coverage does not equate to behavior coverage~\cite{fraser2012behaviourally}. 
Traditional testing methods can not always assure thorough testing. 
Contrarily, LBT operates on the belief that comprehensive testing occurs when every computational action of a program is examined~\cite{weyuker1983assessing}. 
Another LBT perk is its iterative construction of the SUT's approximate model, rendering it scalable from straightforward computational models to larger problems~\cite{meinke2018learning}. 
Performance-wise, Meinke and Niu~\cite{meinke2010learning} found LBT to outperform random testing in speed for error detection in SUTs. 
This efficiency is because LBT's test suite exhibits significantly fewer redundancies. 
Owing to its efficacy, LBT methodologies have found applications in testing procedural~\cite{papadopoulos2015black, fraser2015assessing, walkinshaw2017uncertainty, meinke2011learning} and reactive software~\cite{meinke2011incremental, meinke2013lbtest, 10.1007/978-3-642-24580-0_8}.
\section{Methodology}
\begin{figure}
  \centering
  \includegraphics[width=.9\columnwidth,height=15cm]{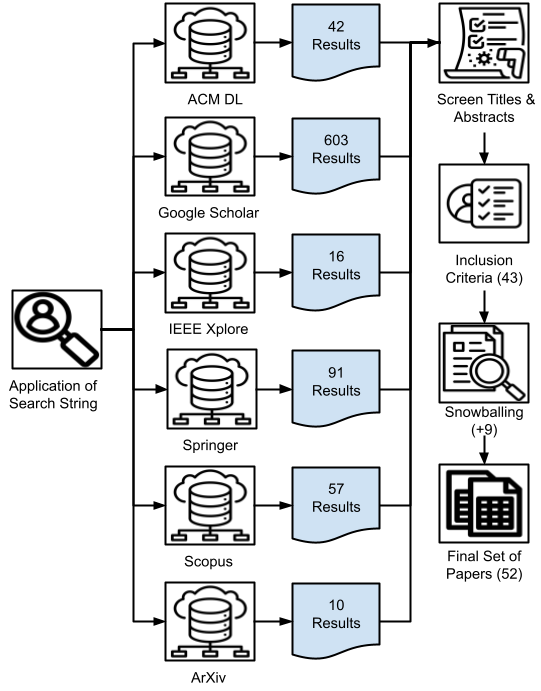}
  \caption{Paper Searching \& Selection Process}
  \label{fig:search_process}
\end{figure}

We have employed a rigorous methodology following the guidelines of Kitchenham~\cite{article4578456} to perform systematic literature reviews in software engineering. 
We utilized a step-by-step approach to collect the most pertinent research papers, leveraging resources such as the ACM Digital Library, Google Scholar, IEEE Xplore, Scopus, Springer and ArXiv. The initial step involved conducting searches on these platforms using a predefined search string (as outlined in table~\ref{tab:protocol_summary}) which served as the basis for identifying relevant studies to be reviewed. To maintain the relevance and quality of the studies incorporated into this review, our inclusion criteria focus on three key aspects during the paper selection process which are also outlined in table~\ref{tab:protocol_summary}. The study selection process, illustrated in Figure~\ref{fig:search_process}, began with an initial search across six sources, resulting in a total of 819 studies, including duplicates.

Given the unexplored nature of this field and the limited number of studies available, we have opted not to impose a strict time limit on the publication date during the search and snowballing process. To maintain consistency and ensure accessibility, papers published in languages other than English were excluded from consideration. To refine the search further, we screened titles and abstracts according to the predefined string (as listed in Table ~\ref{tab:protocol_summary}) and their relevance to the established inclusion criterion. This screening process led to the identification of 43 studies that met the specified criteria.

We also employed a paper snowballing By following the citations and references of the identified paper. We aimed to expand our search and uncover additional studies that may have been missed in our initial search. Initially, we conducted backward snowballing, uncovering further literature by exploring articles referenced in heavily cited papers that elucidate the concept of LBT. After completing backward snowballing, we added more related papers to the existing set. Subsequently, we conducted forward snowballing on the expanded set of papers. 
In forward snowballing, instead of examining the papers a given LBT-related paper has cited, we attempted to find papers that cited the original. We employed these two additional approaches because not all papers related to LBT contain the term ``Learning Based Testing'' or similar expressions in their titles~\cite{walkinshaw2017uncertainty,sharma2022property,papadopoulos2015black}. 
After adding each new paper, we repeated our snowballing approach until no new related paper was found. In total, we collected 52 papers for the final review.

The first three selection criteria mentioned in table~\ref{tab:protocol_summary} are not mutually exclusive, indicating that a paper may fulfill multiple criteria or at least a single criterion to be included in this literature review. 
Some papers were chosen even if they were not directly related to LBT. For instance, Budd and Angluin~\cite{budd1982two} did not specifically propose an LBT method. 
Still, we included the paper because it discusses the equivalent mutant and test set adequacy problem, which inspired LBT. 
As the concept of LBT aims to address both of these issues, we selected this paper to fulfill the first criterion. 
For the third criterion, we chose papers that propose a practical framework or tool based on any theoretical LBT method.
Regarding case studies, we selected papers using LBT methods to test industrial applications or at least simplified versions of those systems. 
Papers discussing testing on simple SUT in an experimental environment were not included for this criterion. 
Table~\ref{table:selection_criteria} summarizes our paper classification based on these criteria.

\begin{table}[t]
  \caption{Protocol Summary}
\begin{tabularx}{\columnwidth}{
    |>{\hsize=0.14\hsize}p{\linewidth}|
    >{\hsize=0.86\hsize}X|
  }
    \hline
    Inclusion Criteria &     
    \begin{enumerate}[leftmargin=*]
        \item The articles which discusses foundational notion of Learning-based testing.
         \newline \textbf{OR}
        \item The article which proposes theoretical frameworks that offer diverse implementations of core LBT concepts, including the proposition of testing tools based on these theoretical foundations.
         \newline \textbf{OR}
        \item The articles which presents case studies of LBT that emphasize innovative implementations of this technique and assess the effectiveness of the proposed approach on the SUT.
        \newline \textbf{AND}
        \item The articles which are written in english.
    \end{enumerate} 
    \\
    \hline
    Search String &
    ``Learning-Based Testing''
    \\
    \hline
  \end{tabularx}

   \label{tab:protocol_summary}
\end{table}

\begin{table*}
 \caption{\label{table:selection_criteria} Papers Based on 
 Selection Criteria}
  \begin{tabularx}{\linewidth}{
    |>{\hsize=0.2\hsize}X|
    >{\hsize=0.8\hsize}X|
  }
    \hline
    \textbf{Category} & \textbf{Papers}
    \\

    \hline
    Conceptual Introduction &  \citet{budd1982two}(\citeyear{budd1982two}),
    \citet{weyuker1983assessing}(\citeyear{weyuker1983assessing}),
    \citet{cherniavsky1987recursion}(\citeyear{cherniavsky1987recursion}),
        \citet{romanik1996using}
       (\citeyear{romanik1996using}),
       \citet{romanik1997approximate}
       (\citeyear{romanik1997approximate}),
       \citet{peled1999black}
       (\citeyear{peled1999black})\\
    \hline
    Theoretical Approaches & 
    \citet{meinke2004automated}(\citeyear{meinke2004automated}), 
    \citet{raffelt2005learnlib}
    (\citeyear{raffelt2005learnlib}),
    \citet{walkinshaw2009iterative}(\citeyear{walkinshaw2009iterative}),
    \citet{meinke2010learning}(\citeyear{meinke2010learning}),
    \citet{meinke2011incremental}(\citeyear{meinke2011incremental}),
    \citet{walkinshaw2011assessing}
    (\citeyear{walkinshaw2011assessing}),
    \citet{10.1007/978-3-642-24580-0_8}
    (\citeyear{10.1007/978-3-642-24580-0_8}),
    \citet{fraser2012behaviourally}
    (\citeyear{fraser2012behaviourally}),
    \citet{choi2013guided}
    (\citeyear{choi2013guided}),
     \citet{fraser2015assessing}
    (\citeyear{fraser2015assessing}),
    \citet{papadopoulos2015black}
    (\citeyear{papadopoulos2015black}),
      \citet{walkinshaw2017uncertainty}
    (\citeyear{walkinshaw2017uncertainty}),
    \citet{zhang2017integration}
    (\citeyear{zhang2017integration}),
    \citet{fiteruau2017learning}
    (\citeyear{fiteruau2017learning}),
    \citet{groz2018revisiting}
    (\citeyear{groz2018revisiting}),
    \citet{weiss2018extracting}
    (\citeyear{weiss2018extracting}),
     \citet{novella2019automatic}
    (\citeyear{novella2019automatic}),
    \citet{aichernig2020passive}
    (\citeyear{aichernig2020passive}),
    \citet{waga2020falsification}
    (\citeyear{waga2020falsification}),
    \citet{mayr2020fly}(\citeyear{mayr2020fly}),
    \citet{shijubo2021efficient}
    (\citeyear{shijubo2021efficient}),
\citet{sharma2021mlcheck},
    (\citeyear{sharma2021mlcheck}),
        \citet{pferscher2021fingerprinting}
    (\citeyear{pferscher2021fingerprinting}),
    \citet{mazhar2021dkl}
    (\citeyear{mazhar2021dkl}),
    \citet{aichernig2021learning}
    (\citeyear{aichernig2021learning}),
    \citet{meinke2021use}(\citeyear{meinke2021use}),
        \citet{sharma2022property}
    (\citeyear{sharma2022property}),
            \citet{quddus2022structural}
    (\citeyear{quddus2022structural}),
    \citet{muskardin2022learning}(\citeyear{muskardin2022learning})
    \\
    \hline
     Frameworks, Tools \& \newline Libraries &  \citet{raffelt2005learnlib}
    (\citeyear{raffelt2005learnlib}),
    \citet{meinke2013lbtest}
    (\citeyear{meinke2013lbtest}),
    \citet{choi2013guided}
    (\citeyear{choi2013guided}),
     \citet{papadopoulos2015black}
    (\citeyear{papadopoulos2015black}),
    \citet{fiteruau2017learning}
    (\citeyear{fiteruau2017learning}),
    \citet{bainczyk2017model}
    (\citeyear{bainczyk2017model}),
        \citet{meinke2017learning}
    (\citeyear{meinke2017learning}),
    \citet{khosrowjerdi2018learning}
    (\citeyear{khosrowjerdi2018learning}),
    \citet{sharma2021mlcheck}
    (\citeyear{sharma2021mlcheck})
    \citet{meinke2021use}(\citeyear{meinke2021use}),
    \citet{muvskardin2022aalpy}(\citeyear{muvskardin2022aalpy})
    \\
    \hline
    Case Studies & 
    \citet{raffelt2009dynamic}(\citeyear{raffelt2009dynamic}),
    \citet{walkinshaw2010increasing}(\citeyear{walkinshaw2010increasing}),
    \citet{meinke2011learning}
    (\citeyear{meinke2011learning}),
    \citet{meinke2011incremental}
    (\citeyear{meinke2011incremental}),
    \citet{steffen2011simplified}
    (\citeyear{steffen2011simplified}),
    \citet{10.1007/978-3-642-24580-0_8}
    (\citeyear{10.1007/978-3-642-24580-0_8}),
    \citet{feng2013case}
    (\citeyear{feng2013case}),
    \citet{choi2013guided}
    (\citeyear{choi2013guided}),
    \citet{meinke2015learning}(\citeyear{meinke2015learning}),
    \citet{sophia2016learning}
    (\citeyear{sophia2016learning}),
    \citet{khosrowjerdi2017automated}
    (\citeyear{khosrowjerdi2017automated}),
    \citet{meinke2017learning}
    (\citeyear{meinke2017learning}),
    \citet{fiteruau2017learning}
    (\citeyear{fiteruau2017learning}),
    \citet{tappler2017model}
    (\citeyear{tappler2017model}),
        \citet{zhang2017integration}
    (\citeyear{zhang2017integration}),
    \citet{bainczyk2017model}
    (\citeyear{bainczyk2017model}),
     \citet{khosrowjerdi2018virtualized}
(\citeyear{khosrowjerdi2018virtualized}),
    \citet{khosrowjerdi2018learning}
    (\citeyear{khosrowjerdi2018learning}),
    \citet{aichernig2019learning}
    (\citeyear{aichernig2019learning}),
    \citet{waga2020falsification}
    (\citeyear{waga2020falsification}),
    \citet{mayr2020fly}(\citeyear{mayr2020fly}),
     \citet{shijubo2021efficient}
    (\citeyear{shijubo2021efficient}),
    \\
    \hline
\end{tabularx}

\end{table*}

\begin{table*}
  \begin{tabularx}{\linewidth}{
    |>{\hsize=0.2\hsize}X|
    >{\hsize=0.8\hsize}X|
  }
    \hline
     &  
    \citet{mayr2020fly}(\citeyear{mayr2020fly}),
     \citet{shijubo2021efficient}
    (\citeyear{shijubo2021efficient}),
    \citet{pferscher2021fingerprinting}
    (\citeyear{pferscher2021fingerprinting}),
     \citet{aichernig2021learning}
    (\citeyear{aichernig2021learning}),
    \citet{meinke2021use}(\citeyear{meinke2021use}),
        \citet{khan2022debugging}
    (\citeyear{khan2022debugging}),
        \citet{quddus2022structural}
    (\citeyear{quddus2022structural}),
    \citet{muvskardin2022aalpy}(\citeyear{muvskardin2022aalpy})
    \\
    \hline
  \end{tabularx}
\end{table*}

We aimed to cover theoretical aspects and existing research, current trends, and case studies to offer a comprehensive perspective on the LBT research field. 
Backward snowballing assisted in identifying papers related to theoretical conceptions and existing research works and forward snowballing helped uncover the current state-of-the-art research in LBT domain.
The first author conducted the initial filtering of papers, carefully documenting each search result and applying the inclusion and exclusion criteria. The second author then replicated the search process and reviewed the results to ensure consistency. Any disagreements between the authors were discussed and resolved collaboratively, ensuring an unbiased final selection. Both authors independently examined the pool of papers to guarantee the accurate extraction of information presented in this paper.
\section{Exploration}
In this section, we discuss various facets of LBT. 
These include the process of model inference, its suitability for testing diverse types of software, and the evaluation of the test suite generated through this approach.

\subsection{\textbf{Conceptual Introduction}}
Considerable attention has been devoted to exploring the relationship between model inference and software testing through LBT. 
Prior to the mid-nineties, this research primarily focused on theoretical aspects. 
Various authors introduced this concept with variations in their approaches. 
For example, Weyuker~\cite{weyuker1983assessing} aimed to infer a general model of behavior for a system from a limited set of observed behaviors and developed a proof-of-concept system that inferred LISP programs based on input/output data. They conveyed that traditional adequacy criteria like statement, branch, or path coverage have limitations as they may not guarantee error detection even if their conditions are met. Though direct application of inference adequacy may be impractical, using it as a guide can help generate more effective test data. By iteratively refining test sets based on the adequacy of the inferred program, testers can better ensure that all relevant aspects of the program's behavior are thoroughly tested. This research is generally considered the pioneer of LBT.

Budd and Angluin~\cite{budd1982two} presented two notions to check the correctness of a program. The first notion is checking the correctness through formal verification, which is exhaustive and inefficient. The second notion is checking the correctness of the program by checking its behavior on only a subset of test cases, which is more pragmatic. In summary, to identify the smallest set of observations necessary to reveal the full spectrum of a system's behavior, Budd and Angluin~\cite{budd1982two} conveyed the theory that if the inferred model is equivalent to the SUT, the test set used to infer the model is adequate. This theory is the basic foundation on which LBT stands.

Cherniavsky and Smith~\cite{cherniavsky1987recursion} introduced a distinctive perspective on LBT rooted in recursion theory by connecting program testing with the principles of inductive inference, resulting in a unique and innovative testing approach. They highlighted the similarities between program testing and inductive inference, emphasizing how both involve deriving programs from finite samples of input/output behavior. The authors demonstrated that performing inference tasks is a more challenging than testing, particularly in the context of recursively enumerable sets of functions. The complexity of synthesis, ambiguity in learning, vast search space, and complexity of program representation were identified as key factors contributing to the greater difficulty of inference.


Romanik and Vitter~\cite{romanik1996using} introduced a testing complexity measure known as Vapnik-Chervonenkis (VC) dimension, where they established minimum and maximum limits for the number of test cases needed to determine if a program is nearly correct or distinguish it from all other programs in the same category. These test cases were presented as pairs of input and output. For example, if we consider any binary classification problem, the VC Dimension is the maximum number of points that can be arranged in any manner such that the hypothesis space can separate them perfectly. In further research, Romanik~\cite{romanik1997approximate} integrated concepts from Probably Approximately Correct (PAC)~\cite{valiant1984theory} learning with software testing to introduce the idea of approximate testing, which says that a finite number of samples ensures that a learned hypothesis is approximately correct with high probability. This leveraged the VC dimension to analyze the testability of different program classes, with a particular emphasis on identifying classes that are challenging or impossible to test using conventional testing approaches. 

Peled et al.~\cite{peled1999black} introduced a method for black box checking to verify the behavior of a system without access to its internal workings. The proposed approach employs Angluin's L* algorithm for learning a finite state machine as a model derived from the reactive system. The strategy involves an iterative process of inferring the system's behavior through learning and refining the inferred model based on testing results - which is the fundamental concept of LBT.

\subsection{\textbf{Model Inference}}
In most of the proposed LBT approaches, researchers employed Active Automata Learning (AAL) to model the behavior of the SUTs. AAL tools are crucial in this process, enabling the modeling of SUT behavior through state machines. For example, 
Raffelt et al.~\cite{raffelt2005learnlib} introduced LearnLib, a Java implementation of a modified version of Angluin’s L* algorithm~\cite{angluin1987learning}, which involves posing membership and equivalence queries to refine a hypothesis automaton, to generate Deterministic Finite Automata (DFA) and Mealy machines for model inference in sequential systems. Similarly, Muvskardin et al.~\cite{muvskardin2022aalpy} proposed a Python library AALpy, offering efficient learning of deterministic, non-deterministic, and stochastic systems. The modular architecture of these AAL tools allows for easy experimentation with new algorithms and oracles, making them adaptable for various research and testing needs.

These AAL tools have been extensively used in various LBT approaches. For example,
Waga~\cite{waga2020falsification} introduced FalCAuN, where the learning process involves utilizing LearnLib to acquire models of Cyber-Physical Systems (CPSs) from their behaviors, which facilitated the extraction of Mealy machines representing CPS models through black-box checking (BBC). During this process, the CPS models were interacted with through inputs, and their corresponding outputs were observed to infer the underlying system behavior. By actively querying the system and analyzing its responses, LearnLib inferred models that approximated the CPS behavior. These models were then used for further analysis, such as equivalence testing and falsification, to verify system properties and identify potential issues. In further research, Shijubo et al.~\cite{shijubo2021efficient} extended FalCAuN by introducing strengthened Linear Temporal Logic (LTL) formulas for model checking. This enhancement aims to improve the efficiency of the BBC process in two ways. 1) By the number of equivalence tests needed, which is expensive, and 2) using model checking with stronger LTL formulas to refine the learned Mealy machine model. Aichernig et al.~\cite{aichernig2021learning} proposed a LearnLib-based LBT approach where the conformance testing of the learned mealy machine of the SUT is done using fuzzing. This is the first LBT approach that integrated fuzzing with LBT.

Walkinshaw et al.~\cite{walkinshaw2009iterative} introduced an iterative refinement technique utilizing heuristics to learn a labeled transition system (LTS) and employed model-based testing to generate tests for conformance checking. 
They demonstrated the feasibility of using passive inference algorithms for active inference. 
The SUT is initially approximated with a small set of test cases in their proposed method. 
Subsequently, new test cases are generated from the inferred state machine. In cases where there is a failing test for the LTS, the method incorporates the failing test sets into the test cases generated from the state machine and learns a new LTS. 
Otherwise, the method returns the LTS.
Meinke and Niu~\cite{10.1007/978-3-642-24580-0_8} employed the complete term rewriting system generator (CGE)~\cite{meinke2010cge} algorithm to learn an extended Mealy automaton (EMA) for inferring the model of a reactive system.
Fiter{\u{a}}u-Bro{\c{s}}tean and Howar~\cite{fiteruau2017learning} proposed a learning-based testing framework that approximates the windowing behavior of the TCP protocol by using SL$^*$ to learn a register automaton. Quddus and Sindhu~\cite{quddus2022structural} proposed an LBT method where the process begins with an initial set of input/output pairs from the SUT, which are used to construct a DFA model of the SUT. Using a model checker, the hypothesis model is then checked against the LTL requirements. If the model violates any LTL requirement, the model checker generates a counterexample, which serves as a new test case. This test case is executed on the SUT to gather more input/output data. If the SUT passes the test case, the hypothesis model is deemed incorrect and refined with the new data. If the SUT fails the test case, indicating that the input/output pair does not satisfy the LTL requirement, LBT terminates as a true negative is identified. This process of generating hypothesis models, checking them against LTL requirements, generating counterexamples, and refining the model continues iteratively, improving the model's accuracy with each iteration.

Pferscher and Aichernig~\cite{pferscher2021fingerprinting} presented an LBT approach to automatically infer a finite-state machine (FSM) of the Bluetooth Low Energy (BLE) protocol implementations in peripheral devices through active automata learning techniques by leveraging an improved variant of the L* algorithm. Groz et al.~\cite{groz2018revisiting} proposed hW-inference method, a novel LBT approach for inferring FSM models from non-resettable black-box systems. This method combines active learning techniques with heuristic methods to iteratively query the system, refine hypotheses of homing sequences and characterization sets, and construct an inferred FSM model.
Novella et al.~\cite{novella2018improving} outlined the methodology for an Extended Labelled Transition System (ELTS) based model inference and testing technique for black-box SUT. In further research, they presented an LBT method for testing the GUI of Android applications that leverage the test results for dynamically learning the ELTS-based model~\cite{novella2019automatic}. Choi et al.~\cite{choi2013guided} proposed  SwiftHand, an LBT approach where the inferred model is based on a finite state machine (FSM) that abstracts the GUI states and transitions of an Android app. This FSM is learned and refined dynamically during the testing process, enabling SwiftHand to generate user inputs that effectively explore new and previously unexplored states of the app.

Meinke, Karl, and Sindhu~\cite{meinke2011incremental} introduced the Incremental Kripke Learning algorithm to model reactive systems. The IKL (Incremental Kripke Learning) algorithm aims to learn a SUT with a deterministic k-bit Kripke structure by incrementally learning individual 1-bit Deterministic Finite Automata (DFA) for each bit. Meinke and Khosrowjerdi~\citeyear{meinke2021use} used this concept and presented ROBOTest, a constrained active machine learning (CAML) architecture, to conduct use-case testing rather than unit testing. Their approach tackles the scalability issues in active automata learning in highly constrained situations. Zhang et al.~\cite{zhang2017integration} utilized a finite automaton model to prevent reactive black-box systems from reaching faulty sections. 
Their research integrated LBT with Supervisory Control Theory (SCT) to ensure the safe usage of black-box reactive systems.

Walkinshaw~\cite{walkinshaw2011assessing} proposed using the PAC framework for empirically assessing the adequacy of test sets for black-box systems using Version Spaces and the VC Dimension. Fraser and Walkinshaw~\cite{fraser2012behaviourally} used this concept and proposed BESTEST, where PAC framework is integrated into the search-based testing technique so that the generated test sets are not only comprehensive in terms of code coverage but also adequately reflect the software's intended behavior. The genetic algorithm optimizes for both these criteria, guided by the PAC principles, while maintaining the efficiency and independence of the test sets.
One issue with the LBT utilizing automata learning-based inference lies in its inherent assumption that the SUT follows a sequential pattern, which led to the necessity of adopting a data-driven approach to conduct model inference in non-sequential SUT scenarios.

Researchers have applied various methods for model inference in the context of LBT. 
The Model-Inference driven Testing (MINTEST) framework, proposed by Papadopoulos and Walkinshaw~\cite{papadopoulos2015black}, leverages WEKA’s J48~\cite{hall2009weka} implementation of Quinlan’s C4.5 algorithm~\cite{quinlan2014c4} to infer decision trees from program executions. 
Subsequently, it employs the Z3 solver~\cite{de2008z3} to generate and execute tests based on these trees. 
The results are analyzed to generate new test inputs, and this cycle continues. 
Sharma et al.~\cite{sharma2021mlcheck} proposed MLCheck, which enables property-driven testing for machine learning models by allowing developers to specify properties they want to validate and then generate test cases to check if these properties are satisfied. It trains a white-box machine learning model approximating the black-box model under test, translates specified properties and the white-box model into logical formulas, and uses an SMT solver to check satisfiability. Counterexamples are extracted and added to the test suite, and the white-box model is retrained iteratively to improve approximation quality. This systematic approach ensures quality and reliability across various application areas. MLCheck primarily utilizes two types of models as white-box approximations of the black-box machine learning models under test: decision trees and neural networks. 

In further research, Sharma et al.~\cite{sharma2022property} used this concept and presented an iterative process for testing numerical functions against user-defined properties, where they employed MLCheck to test black-box numerical functions using ML model of the SUT, instead of using automata learning. The distinction in their approach, compared to the proposed method by Papadopoulos et al.~\cite{papadopoulos2015black}, lies in their use of the tree to generate counterexamples of the property under test. In contrast, Papadopoulos et al.~\cite{papadopoulos2015black} translated the decision tree to logic and utilized Z3 to generate test inputs covering the branches. Fraser and Walkinshaw~\cite{fraser2015assessing} have employed various machine learning algorithms, including C4.5 Decision Tree, Naive Bayesian Network, and AdaBoost, among others, for model inference to tackle the limitation of their previous approach which used the PAC framework~\cite{fraser2012behaviourally}. 
The authors propose using k-fold cross-validation (CV) to quantify behavioural adequacy of the model instead of relying solely on the PAC framework. 
More specifically, They used ML algorithms to solve the test set adequacy problem with LBT, which focuses on both effective test suite generation and observable program behavior coverage.

Researchers have not limited themselves to using only automata and machine learning algorithms as inference techniques in LBT. 
They have also explored the utilization of genetic algorithms as model inference techniques~\cite{meinke2018learning}. 
Walkinshaw and Fraser~\cite{walkinshaw2017uncertainty} proposed the Testing By Committee algorithm, an LBT approach based on an active learning algorithm known as Query By Committee. 
This method actively employs a genetic programming inference Engine to generate a set of models for model inference in each iteration. Aichernig et al.~\cite{aichernig2020passive} also used genetic programming to build a timed automaton of the SUT. They proposed an iterative method to improve the model, a similar approach to what Walkinshaw et al.~\cite{walkinshaw2009iterative} proposed using a state machine. However, in the new approach, the test traces are generated with random walks, and only counterexamples are used to create the hypothesis model using genetic programming.

Meinke~\cite{meinke2004automated} proposed a technique to use polynomial models to approximate the SUT. In their research, they proposed a set of piecewise overlapping polynomial models instead of a single global model of a 1-dimensional numerical program. 
Meinke and Niu~\cite{meinke2010learning} further extended this exploration by using n-dimensional polynomial equations for model inference that can support n-wise testing. They used the model to infer a high dimensional numerical program and generated test cases by applying a satisfiability algorithm to the model. 

LBT approaches have recently focused on testing deep learning models as well. For instance, Mayr et al.~\cite{mayr2020fly} proposed an algorithm called Bounded-L$^*$, which constructs a DFA representing the language of a Recurrent Neural Network (RNN). This DFA serves as an approximation of the RNN's behavior concerning the specified property. Weiss et al.~\cite{weiss2018extracting} developed an algorithm to answer equivalence queries between the RNN and a candidate automaton and uses this to learn a minimal DFA that captures the behavior of the RNN. Aichernig et al.~\cite{aichernig2024learning} proposed an LBT approach where behavioral model of an RNN is extraced using state machine, they used constrained training technique to generate RNN, the behavioral model of the SUT. Muskardin et al.~\cite{muskardin2022learning} implemented the learning algorithms and equivalence oracles for active automata learning of RNN models using AALpy~\cite{muvskardin2022aalpy}.

Table~\ref{table:inference_techniques} summarizes different inference techniques used in LBT.
\begin{table}
    \caption{\label{table:inference_techniques} Model Inference Techniques in LBT}
  \begin{tabularx}{\linewidth}{
    |>{\hsize=0.16\hsize}X|
    >{\hsize=0.84\hsize}X|
  }
    \hline
    \textbf{Model Type} & \textbf{Papers} \\
    \hline
    Automata \newline Learning & 
    \citet{raffelt2005learnlib}
    (\citeyear{raffelt2005learnlib}), 
    \citet{shahbaz2009inferring}
    (\citeyear{shahbaz2009inferring}),
    \citet{walkinshaw2009iterative}
    (\citeyear{walkinshaw2009iterative}),
    \citet{10.1007/978-3-642-24580-0_8}
    (\citeyear{10.1007/978-3-642-24580-0_8}),
    \citet{meinke2011incremental}
    (\citeyear{meinke2011incremental}),
    \citet{choi2013guided}
    (\citeyear{choi2013guided}),
    \citet{zhang2017integration}(\citeyear{zhang2017integration}),
    \citet{zhang2017integration}
    (\citeyear{zhang2017integration}),
    \citet{fiteruau2017learning}
    (\citeyear{fiteruau2017learning}),
    \citet{groz2018revisiting}
    (\citeyear{groz2018revisiting}),
     \citet{novella2019automatic}
    (\citeyear{novella2019automatic}),
     \citet{pferscher2021fingerprinting}
    (\citeyear{pferscher2021fingerprinting}),
    \citet{mazhar2021dkl}
    (\citeyear{mazhar2021dkl}),
     \citet{quddus2022structural}
    (\citeyear{quddus2022structural}),
    \citet{muskardin2022learning}
    (\citeyear{muskardin2022learning}),
    \citet{meinke2011incremental} (\citeyear{meinke2011incremental}), \citet{meinke2013lbtest} (\citeyear{meinke2013lbtest}),
    \citet{khosrowjerdi2018learning}
    (\citeyear{khosrowjerdi2018learning}),
    \citet{mazhar2021dkl}
    (\citeyear{mazhar2021dkl}), 
    \citet{muvskardin2022aalpy}(\citeyear{muvskardin2022aalpy}),
    \citet{muskardin2022learning}
(\citeyear{muskardin2022learning}) 
    \\
    \hline
    Machine Learning & 
    \citet{papadopoulos2015black}
    (\citeyear{papadopoulos2015black}),
    \citet{fraser2015assessing}
    (\citeyear{fraser2015assessing}),
    \citet{sharma2021mlcheck}
    (\citeyear{sharma2021mlcheck}),
    \citet{sharma2022property}
    (\citeyear{sharma2022property}),
    \citet{aichernig2024learning}
    (\citeyear{aichernig2024learning})
    \\
    \hline
    Genetic \newline Inference & 
    \citet{ghani2008strengthening}(\citeyear{ghani2008strengthening}), 
    \citet{fraser2012behaviourally}(\citeyear{fraser2012behaviourally}), 
    \citet{walkinshaw2017uncertainty}(\citeyear{walkinshaw2017uncertainty}), 
    \citet{aichernig2020passive}(\citeyear{aichernig2020passive})
    \\
    \hline
    Polynomial Model & \citet{meinke2004automated} (\citeyear{meinke2004automated}), \citet{meinke2010learning} (\citeyear{meinke2010learning})
    \\
    \hline
  \end{tabularx}
\end{table}

\begin{table*}
 \caption{\label{table:paper_summary} Summary of The Approaches and Frameworks}
  \begin{tabularx}{\linewidth}{|>{\hsize=0.03\hsize}X|>{\hsize=0.19\hsize}X|>{\hsize=0.08\hsize}X|>{\hsize=0.52\hsize}X|>{\hsize=0.22\hsize}X|}
    \hline
    \textbf{\#} & \textbf{Ref} & \textbf{Year} & \textbf{Summary}
    & \textbf{Evaluated On}
     \\
    \hline
    \rownumber &
    \citet{budd1982two} &
    \citeyear{budd1982two} &
    Presented the idea that if an inferred model is equivalent to a certian SUT, the test set used to infer the model is adequate for testing the SUT.
    &
    N/A
    \\ \hline
    \rownumber &
    \citet{weyuker1983assessing}
    & \citeyear{weyuker1983assessing}
    & Proposes the use of inference adequacy as a criterion for test data adequacy that tackles both of the issues related to testing mentioned in~\cite{budd1982two}
    & programs written in PL/I from~\cite{summers1977methodology}
    \\ \hline
    \rownumber & \citet{cherniavsky1987recursion}
    & \citeyear{cherniavsky1987recursion}
    & Introduced the notion of recursion-theoretic perspective to analyze program testing, emphasizing the incomparability between testing and inference.
    & N/A
    \\ \hline
    \rownumber &
    \citet{romanik1997approximate} &
    \citeyear{romanik1997approximate} &
    Integrated the concept of PAC~\cite{valiant1984theory} with software testing.
    & N/A
    \\ \hline
    \rownumber & 
    \citet{peled1999black}
    & \citeyear{peled1999black}
    & Introduced Black-box checking method using Angluin's L* algorithm.
    & N/A
    \\ \hline
    \rownumber &
    \citet{meinke2004automated}
    & \citeyear{meinke2004automated}
    & Used piecewise overlapping polynomial models for SUT approximation
   & Simple numerical functions    
    \\
    \hline
    \rownumber &
    \citet{raffelt2005learnlib}
    &
    \citeyear{raffelt2005learnlib}
    & Proposed LearnLib
    & Web application and telephone hardware
    \\
    \hline
    \rownumber & \citet{raffelt2009dynamic}
    & \citeyear{raffelt2009dynamic} &
    Conducted a case study on the effectiveness of LearnLib~\cite{raffelt2005learnlib}
    & Mantis Bug Tracking System, Java Router
    \\ \hline
    \rownumber &
    \citet{walkinshaw2009iterative}
    &\citeyear{walkinshaw2009iterative}
    & Proposed LBT based on inferring a reverse-engineering state machine of the SUT
    & Erlang implementation of a FTP client
    \\
    \hline
    \end{tabularx}
\end{table*}

\begin{table}
  \begin{tabularx}{\linewidth}{|>{\hsize=0.03\hsize}X|>{\hsize=0.19\hsize}X|>{\hsize=0.08\hsize}X|>{\hsize=0.52\hsize}X|>{\hsize=0.22\hsize}X|}
    \hline
    \textbf{\#} & \textbf{Ref} & \textbf{Year} & \textbf{Summary}
    & \textbf{Evaluated On}
     \\
    \hline
    \rownumber &
        \citet{meinke2010learning}
    &
    \citeyear{meinke2010learning}
    & 
    Extended the concept from~\cite{meinke2004automated} by using n-dimensional polynomial equations for model inference that can support n-wise testing. 
    &
    Randomly Generated numerical functions
    \\ \hline
    \rownumber & \citet{walkinshaw2010increasing} 
    &\citeyear{walkinshaw2010increasing}
    & Conducted a case study based on the LBT proposed in~\cite{walkinshaw2009iterative}
    & Linux TCP/IP stack
    \\
    \hline
     \rownumber &
    \citet{meinke2011incremental}
    &\citeyear{meinke2011incremental}
    & Presented IKL algorithm.
    & 8 state cruise controller and a 38 state 3-floor elevator model
    \\
    \hline
    \rownumber &
    \citet{walkinshaw2011assessing}
    &\citeyear{walkinshaw2011assessing}
    &
    Introduced a PAC based framework for black-box systems testing by LBT.
    &
    SSH client simulator 
    \\ \hline
    \rownumber & \citet{10.1007/978-3-642-24580-0_8}
    &
    \citeyear{10.1007/978-3-642-24580-0_8}
    & Introduced CGE algorithm and integrated term rewriting with LBT
    & TCP/IP protocol
    \\
    \hline
    \rownumber
    & \citet{meinke2011learning}
    &\citeyear{meinke2011learning}
    & Presented a comparison study between LBT approaches proposed in~\cite{meinke2010learning,meinke2011incremental,10.1007/978-3-642-24580-0_8}
    & Elevator control program, TCP/IP protocol, Multidimensional piecewise continuous functions.
    \\ \hline
    \rownumber &
    \citet{steffen2011simplified}
    &\citeyear{steffen2011simplified}
    & Presented a case study where LearnLib~\cite{raffelt2005learnlib} is used to test OCS
    & OCS
    \\ \hline
         \rownumber &
    \citet{fraser2012behaviourally}
    &\citeyear{fraser2012behaviourally}
    & Proposed BESTEST based on the concept of the PAC framework~\cite{walkinshaw2011assessing}
    &
    Simple numerical functions.
    \\
    \hline
\end{tabularx}
\end{table}

\begin{table}
  \begin{tabularx}{\linewidth}{|>{\hsize=0.03\hsize}X|>{\hsize=0.21\hsize}X|>{\hsize=0.08\hsize}X|>{\hsize=0.52\hsize}X|>{\hsize=0.22\hsize}X|}
    \hline
    \textbf{\#} & \textbf{Ref} & \textbf{Year} & \textbf{Summary}
    & \textbf{Evaluated On}
     \\
    \hline

    \rownumber &
    \citet{choi2013guided}
    &
    \citeyear{choi2013guided}
    & proposed  SwiftHand to test GUI of andriod apps.
    & 10 apps from F-Droid open app market.
    \\
    \hline
    \rownumber &
    \citet{meinke2013lbtest}
    &
    \citeyear{meinke2013lbtest} &
    Presents LBT tool LBTest based on the proposed IKL algorithm~\cite{meinke2011incremental}.
    &
    Cruise Controller Application
    \\
    \hline
    \rownumber
    & \citet{feng2013case}
    &
    \citeyear{feng2013case}
    & Conducted a case study using LBTest~\cite{meinke2013lbtest} to test different commercial softwares
    & FAS, BBW, ABS
    
    \\ \hline
    \rownumber &
    \citet{fraser2015assessing}
    &\citeyear{fraser2015assessing}
    & Used ML with Cross Validation to tackle the limitations of BESTEST~\cite{fraser2012behaviourally}.
    & Simple numerical functions.
    \\
    \hline
    \rownumber & \citet{papadopoulos2015black}
    &\citeyear{papadopoulos2015black}
    & Proposed Decision tree based LBT framework MINTEST &
    Simple numerical functions
    \\ \hline
    \rownumber
    &
    \citet{meinke2015learning}
    & \citeyear{meinke2015learning}
    & Conducted a case study on testing triCalculate, a counter-party credit risk analysis system using LBTest~\cite{meinke2013lbtest}
    & triCalculate
    \\ \hline    
    \rownumber & \citet{sophia2016learning} & \citeyear{sophia2016learning} & Conducted a case study on LBTest on ECUs & ECUs
    \\ \hline
        \rownumber &
    \citet{walkinshaw2017uncertainty}
    &\citeyear{walkinshaw2017uncertainty}
    & Proposed a QBC algorithm based on genetic inference for SUT approximation.
    &
    Simple numerical functions
    \\
    \hline
    \rownumber &
    \citet{zhang2017integration}
    &\citeyear{zhang2017integration}
    &
    Integrated the concept of SCT with LBT for reactive systems using LBTest~\cite{meinke2013lbtest}
    &
    BBW controller system
    \\
    \hline
    \rownumber & \citet{fiteruau2017learning}
    &\citeyear{fiteruau2017learning} &
    Proposed SL$^*$ algorithm based LBT to test the windowing behavior of the TCP protocol.
    & TCP protocol.
    \\
    \hline
     \rownumber & \citet{khosrowjerdi2017automated} & \citeyear{khosrowjerdi2017automated} & Conducted an industrial case study on ECUs using LBTest
    & BBW, ESTA, DCS, FLD
    \\ \hline
\end{tabularx}
\end{table}
\begin{table}
  \begin{tabularx}{\linewidth}{|>{\hsize=0.03\hsize}X|>{\hsize=0.20\hsize}X|>{\hsize=0.08\hsize}X|>{\hsize=0.52\hsize}X|>{\hsize=0.20\hsize}X|}
    \hline
    \textbf{\#} & \textbf{Ref} & \textbf{Year} & \textbf{Summary}
    & \textbf{Evaluated On}
     \\
    \hline
        \rownumber & \citet{bainczyk2017model} &
    \citeyear{bainczyk2017model} &
    Presented the ALEX tool, a graphical interface to LearnLib~\cite{raffelt2005learnlib} for testing web applications and HTTP-based web APIs
    & 27 TodoMVC applications
    \\ \hline
    \rownumber & \citet{meinke2017learning} & \citeyear{meinke2017learning} & Proposed LBTest 3.x, a multi-core version of LBTest~\cite{meinke2013lbtest} for concurrent learning of the SUTs & Vehicle platooning simulator
    \\ \hline
    \rownumber & \citet{tappler2017model} & \citeyear{tappler2017model} &
    Extended LearnLib~\cite{raffelt2005learnlib} to conduct case study on MQTT brokers
    &
    MQTT brokers
    \\ \hline
        \rownumber & \citet{khosrowjerdi2018learning}
    &\citeyear{khosrowjerdi2018learning}
    & Conducted a case study using LBTest 3.x~\cite{meinke2017learning} &
    Vehicle platooning simulator
    \\ \hline
    
    \rownumber & \citet{groz2018revisiting}
    &\citeyear{groz2018revisiting}
    & Proposed hW-inference, an LBT approach for testing non-resettable black-box systems. 
    & N/A
    \\
    \hline
    \rownumber & \citet{weiss2018extracting} & \citeyear{weiss2018extracting} &
    Proposed an L$^*$ based algorithm to to learn a minimal DFA that captures the behavior of the RNN
    & RNN models trained on Tomita Grammars
    \\ \hline
    \rownumber & \citet{khosrowjerdi2018virtualized} & \citeyear{khosrowjerdi2018virtualized} & proposed FI testing case study using LBTest~\cite{meinke2013lbtest} & ECU applications from Scania CV
    \\ \hline
    \rownumber &
    \citet{novella2019automatic}
    &\citeyear{novella2019automatic}
    & Proposed an LBT based on the concept of ELTS model learning proposed in~\cite{novella2018improving} 
    & A Set of Android applications
    \\ \hline
    \rownumber & \citet{aichernig2019learning}
    & \citeyear{aichernig2019learning}
    & Conducted a Case study on LearnLib~\cite{raffelt2005learnlib} & AVL489 exhaust measurement device
    \\ \hline
    \rownumber & \citet{waga2020falsification}
    &\citeyear{waga2020falsification}
    & Introduced FalCAuN where optimization-based falsification and BBC is combined to present robustness-guided BBC in LBT
    &
     Simulink automatic transmission system model
    \\
    \hline
\end{tabularx}
\end{table}
\begin{table}
  \begin{tabularx}{\linewidth}{|>{\hsize=0.03\hsize}X|>{\hsize=0.20\hsize}X|>{\hsize=0.08\hsize}X|>{\hsize=0.52\hsize}X|>{\hsize=0.20\hsize}X|}
    \hline
    \textbf{\#} & \textbf{Ref} & \textbf{Year} & \textbf{Summary}
    & \textbf{Evaluated On}
     \\
    \hline

        \rownumber & \citet{mayr2020fly} & \citeyear{mayr2020fly} &
    Proposed Bounded-L$^*$ algorithm for effective and efficient verification of properties of RNNs.
    & CCS, HDFS, TATA-box
     \\ \hline
    \rownumber
    &
    \citet{aichernig2020passive}
    &\citeyear{aichernig2020passive}
    & proposed iterative refinement based LBT similar to~\cite{walkinshaw2009iterative}, but used genetic inference for model creation
    & Timed Automaton
    \\ \hline
    
        \rownumber & \citet{shijubo2021efficient}
    &\citeyear{shijubo2021efficient}
    & Enhanced FalCAuN~\cite{waga2020falsification} by introducing LTL formulas for model checking
    & Simulink automatic transmission system model
    \\
    \hline
    \rownumber &
    \citet{mazhar2021dkl}
    &
    \citeyear{mazhar2021dkl}
    & Proposed DKL to solve the  state-space explosion problem of IKL~\cite{meinke2011incremental}
    &  Random deterministic Kripke structures
    \\
    \hline
    \rownumber
    & \citet{sharma2021mlcheck}
    &
    \citeyear{sharma2021mlcheck}
    & Proposed MLCheck for testing ML models.
    &
    \\ \hline
    \rownumber & \citet{aichernig2021learning} &
    \citeyear{aichernig2021learning} &
    Introduced Fuzzing for conformance checking in LBT
    & MQTT protocol
     \\ \hline
    \rownumber & \citet{meinke2021use} & \citeyear{meinke2021use} &
        Proposed a constrained active ML architecture built on LBTest~\cite{meinke2013lbtest}
        & ASM vehicle simulator
    \\ \hline
    \rownumber & \citet{pferscher2021fingerprinting} & \citeyear{pferscher2021fingerprinting} & Used an improved version of L* algorithm with LearnLib to test BLE protocol
    & BLE protocol
    \\ \hline
        \rownumber
    & \citet{sharma2022property}
    &\citeyear{sharma2022property}
    & Used MLCheck~\cite{sharma2021mlcheck} as LBT tool for black-box systems.
    & Aggregation functions
    \\ \hline
    \rownumber & \citet{khan2022debugging}
    &
    \citeyear{khan2022debugging}
    & Presented a comparison study between LBT framework for reactive systems proposed in~\cite{meinke2011incremental} and other MBT methods
    & CCS, ATM
    \\ \hline
\end{tabularx}
\end{table}
\begin{table}
  \begin{tabularx}{\linewidth}{|>{\hsize=0.03\hsize}X|>{\hsize=0.20\hsize}X|>{\hsize=0.08\hsize}X|>{\hsize=0.52\hsize}X|>{\hsize=0.20\hsize}X|}
    \hline
    \textbf{\#} & \textbf{Ref} & \textbf{Year} & \textbf{Summary}
    & \textbf{Evaluated On}
     \\
    \hline
     \rownumber
    &
    \citet{quddus2022structural}
    &\citeyear{quddus2022structural}
    &
    Evaluated the structural coverage of LTL requirements achieved by LBT test suites.
    & CCS, ATM
    \\ \hline
    
    \rownumber & \citet{muvskardin2022aalpy} & \citeyear{muvskardin2022aalpy}
    & proposed AALpy, an active automata learning library implemented in Python. & 
    MQTT, BLE protocol, Vim etc.
    \\ \hline
    \rownumber & \citet{muskardin2022learning}
    & \citeyear{muskardin2022learning}
    & proposed a method for RNN model verification by coverage-guided conformance testing using AALpy~\cite{muvskardin2022aalpy}.
    & RNN models.
    \\ \hline
    \rownumber & \citet{aichernig2024learning} & \citeyear{aichernig2024learning} &
    Proposed an LBT method where RNN model will be inferrred of the SUT using active learning
    & BLE protocol, Tomita Grammars
    \\ \hline
      \end{tabularx}
\end{table}




\subsection{\textbf{Model Checking}}
Model checking generates queries, which are counterexamples within the learned model that challenge the correctness of the system requirements. 
It distinguishes each iterative model from the previous one by producing and checking counterexamples for which the previous and current models in the iteration yield different outputs. 
Consequently, model checking serves as a stopping criterion for LBT~\cite{meinke2011learning}. 
When it can no longer generate counterexamples for the inferred model, it indicates that the system has been adequately tested and the model has converged to an approximate representation of the SUT. In addition, it works as an effective test case generator as well because the counterexamples on the model have a higher probability of resulting in bugs when executed on the SUT.

Furthermore, researchers emphasize that the effectiveness of LBT relies heavily on the efficiency and efficacy of the chosen model checker~\cite{khosrowjerdi2018learning}. 
Various implementations of model checkers have been developed to accommodate different approaches within the domain of LBT. The type of model checkers depends largely on the model inference method that the LBT approach has incorporated.
For example, Meinke and Niu~\cite{meinke2010learning} utilized the Hoon-Collins cylindric algebraic decomposition (CAD) algorithm to perform model checking on the polynomial model, an abstraction of the SUT. The choice is reasonable because the CAD algorithm is explicitly designed to address systems of polynomial equations. SMT solvers have also been used in LBT for model checking. Papadopoulos and Walkinshaw~\cite{papadopoulos2015black} used Z3 solver~\cite{de2008z3} for their Decision Tree based model inference technique. In each iteration, they derive the constraints from the generated tree and each leaf node's path from the root node to generate test cases by the Z3 solver. Then, the output for the test case created by the Z3 solver is checked against the expected output. As already discussed, Sharma et al.~\cite{sharma2022property} have also used the Z3 solver as the model checker for their approach.
In a separate approach, Meinke and Sindhu~\cite{meinke2011incremental} integrated the NuSMV model checker into their tool named LBTest~\cite{meinke2013lbtest}. The rationale behind this choice is that NuSMV supports satisfiability analysis of Kripke structures, which they have used to model reactive systems regarding both LTL and Computation Tree Logic (CTL). \citet{fraser2015assessing} used k-fold cross-validation for their ML-based LBT approach. Quddus and Sindhu~\cite{quddus2022structural}, in their proposed LBT method, have also used NuSMV to get the inferred model checked against the LTL requirements using a model checker.

In their subsequent research, Khosrowjerdi and Meinke~\cite{khosrowjerdi2018learning} employed NuXmv~\cite{cavada2014nuxmv} as the model checker, an extended version of NuSMV, which provides support for \textit{integer} and \textit{real} data type. It supports first-order LTL in conjunction with LBTest 3.x~\cite{khosrowjerdi2018learning}, an improved version of LBTest, to address the spatiotemporal requirements of the SUT. 
Meinke and Niu~\cite{10.1007/978-3-642-24580-0_8} applied a first-order disunification algorithm using basic narrowing as the model checker for their CGE incremental learning algorithm. 
This algorithm infers Extended Mealy Automata (EMA), which can be seen as a Mealy machine over abstract data types (ADT) as inputs and outputs. 
Fraser and Walkinshaw~\cite{fraser2012behaviourally} utilized the PAC framework to assess the adequacy of the model. 
However, PAC relies on the assumption of having two large samples chosen under identical conditions. 
In real-world testing scenarios, there is often a shortage of test cases, and dividing this limited sample into training and test subsets can result in significantly different feature sets, which challenges the reliability of the framework~\cite{fraser2015assessing}.

To address this issue, Fraser and Walkinshaw~\cite{fraser2015assessing} replaced PAC with k-fold cross-validation in their subsequent work, providing a more standardized way to evaluate the inferred model. 
Walkinshaw et al.~\cite{walkinshaw2009iterative} incorporated QuickCheck, an automated testing tool for Erlang, to generate counterexamples in their proposed LBT method.
QuickCheck~\cite{claessen2000quickcheck} has the capability to generate the necessary input sequence to test any path of a given state-machine model.
Even model checking has been done manually as well in the LBT framework. For instance, Fiter{\u{a}}u-Bro{\c{s}}tean and Howar~\cite{fiteruau2017learning} used manual checking of the generated register automata to check whether the model has generated any counter-example or not.



\subsection{\textbf{Reactive System Testing Approaches}}
In addition to its application in procedural programming, LBT has also proven valuable in testing reactive systems. 
Meinke and Sindhu~\cite{meinke2011incremental} employed Kripke structures to model reactive SUTs, along with temporal logic formulas to represent user requirements for these reactive SUTs.
LBTest~\cite{meinke2013lbtest} stands out as a widely used tool for testing reactive systems, building upon this conceptual foundation. 

In further research, Meinke~\cite{meinke2017learning} introduced the multi-core version of LBTest, an enhancement of the original LBTest, enabling concurrent execution of multiple instances of SUTs on a multi-core platform to reduce test latency. 
Another approach discussed earlier by Meinke and Niu~\cite{10.1007/978-3-642-24580-0_8} focuses on testing reactive systems. 
This method is grounded in term rewriting technology, and the authors applied it to test the TIP/IP protocol, demonstrating superior performance compared to random testing.

Aichernig et al.~\cite{aichernig2024learning} introduced a novel machine learning technique to learn minimal finite-state models that represent a reactive system, specifically Mealy machines, by leveraging a specialized RNN architecture and a constrained training method. Muskardin et al.~\cite{muskardin2022learning} proposed an LBT method where AALpy is used to learn RNN models. \citet{quddus2022structural} proposed an LBT framework where trap properties are derived from LTL requirements to capture structural coverage criteria of reactive systems. Pferscher and Aichernig~\cite{pferscher2021fingerprinting} presented active automata learning techniques by leveraging an improved variant of the L* algorithm for testing reactive systems. \citet{aichernig2021learning} proposed using fuzzing in LBT to test MQTT protocols. Zhang et al.~\cite{zhang2017integration} proposed the use of SCT and LBT to ensure the safe reuse of black-box reactive components even when internal modifications are impossible. In this approach, Requirements are expressed in Probabilistic LTL and tested one by one using LBTest. \citet{tappler2017model} integrated LearnLib with a custom mapper component specifically designed for handling the unique characteristics of reactive systems such as MQTT brokers. They integrated MQTT-specific components, such as adapters for communication tasks and client-interface components, with LearnLib to ensure smooth communication between the MQTT environment during the learning process.

\subsection{\textbf{Tools, Libraries \& Frameworks}}
While most approaches in LBT have remained theoretical and confined to experimental environments, there are practical LBT tools and libraries available for real-world testing. 
For instance, in the MINTEST framework~\cite{papadopoulos2015black}, researchers use the WEKA inference framework~\cite{hall2009weka} for model inference. 
To initiate the test generation process, users need to provide a JSON file specifying the SUT, which is then processed with the aid of the Z3 solver~\cite{de2008z3}. 
Walkinshaw~\cite{fraser2012behaviourally} presented BESTEST, an LBT tool where PAC framework~\cite{walkinshaw2011assessing} is integrated into the search-based testing technique.

LBTest~\cite{meinke2013lbtest}, a widely adopted LBT tool utilized in testing numerous reactive and embedded systems, offers a well-organized Graphical User Interface (GUI). This framework is based on a previous LBT approach~\cite{meinke2011incremental}, which used an incremental learning algorithm to learn a DFA model of SUTs, which can be interpreted as a Boolean Kripke structure. This interface facilitates the input of Propositional Linear Temporal Logic (PLTL), SUT interface details, and the execution of tests. Furthermore, researchers have introduced an enhanced iteration of LBTest, known as LBTest 3.x. This iteration allows for the concurrent learning of the SUT, making use of multi-core hardware~\cite{khosrowjerdi2018learning}, which leads to a substantial reduces the required time for the SUT's model inference~\cite{meinke2017learning}. 

Meinke and Khosrowjerdi~\cite{meinke2021use} implemented
ROBOTest, a CAML architecture on top of LBTest to conduct use case testing on the ASM vehicle simulator.
MLCheck~\cite{sharma2021mlcheck} is a tool designed for LBT of machine learning models. It offers developers a systematic approach to validate whether specified requirements are fulfilled by machine learning components in software applications. The tool addresses the growing need for quality assurance in machine learning applications, where ensuring the reliability, fairness, and robustness of models is crucial.

FalCAuN~\cite{waga2020falsification}, a tool for robustness-guided Black-Box Checking (BBC) of Cyber-Physical Systems (CPSs), implemented in Java with LearnLib~\cite{raffelt2005learnlib} for active automata learning and model checking. LearnLib facilitated active automata learning, while model checking was performed using techniques integrated into FalCAuN. 

Bainczyk et al.~\cite{bainczyk2017model} proposed ALEX tool, which serves as a graphical user interface to LearnLib, facilitating the inference of Mealy machines for web applications and HTTP-based web APIs.

Muvskardin et al.~\cite{muvskardin2022aalpy} proposed AALpy, an active automata learning library implemented in Python. It efficiently learns deterministic, non deterministic, and stochastic systems, providing a range of equivalence oracles, learning algorithms, and visualization tools.

\subsection{\textbf{Case Studies}}

Researchers have performed numerous case studies to evaluate the effectiveness of LBT tools in testing real-world industrial systems, including networking and communication protocols, autonomous driving systems, and commercial software across various industries.
 
Meinke et al.~\cite{10.1007/978-3-642-24580-0_8} conducted a case study to test a simplified model of the Transmission Control Protocol (TCP) protocol using their proposed term rewriting and narrowing-based LBT approach. They showed that though their approach is slightly slower than random testing, the proposed LBT method always finds errors with significantly fewer test cases. Walkinshaw et al.~\cite{walkinshaw2009iterative} utilized their reverse-engineering-based model to test an FTP client program based on the Erlang programming language with the assistance of the QuickCheck testing framework. 
In a subsequent case study, Walkinshaw et al.~\cite{walkinshaw2010increasing} generated a test set for the Linux TCP/IP stack using the previously proposed method~\cite{walkinshaw2009iterative}. They compared their inductive testing method with non-inductive methods and showed that the proposed method achieves better coverage than the non-inductive testing techniques. The proposed LBT framework by Fiter{\u{a}}u-Bro{\c{s}}tean and Howar~\cite{fiteruau2017learning} aided in finding confirmed violations of TCP specifications in both Linux and Windows implementation. They tested the windowing property on TCP in both Linux and Windows implementation and revealed a confirmed violation of the RFC 793 standard~\cite{postel1981transmission} in both Linux and Windows. 

Tappler et al.~\cite{tappler2017model} demonstrated the effectiveness of LearnbLib for detecting faults in reactive systems, which is the Message Queuing Telemetry Transport (MQTT) protocol. They found 17 bugs across the implementations, with non-deterministic behavior posing a challenge. The study identified violations of the MQTT protocol specification and discussed the manual effort required for analysis. Efficiency issues were noted, with long runtimes observed, especially in experiments with ActiveMQ and VerneMQ. While the approach effectively detected faults, challenges like non-determinism and efficiency call for further optimization. Pferscher and Aichernig~\cite{pferscher2021fingerprinting} presented a case study on their proposed LBT method of the Bluetooth Low Energy (BLE) protocol, aiming to evaluate the practical application of the approach. The study evaluates the learning framework on five BLE devices, demonstrating the feasibility of active learning in a practical timeframe. The learned models reveal significant variations in BLE stack implementations across devices, confirming the hypothesis that active automata learning enables the fingerprinting of black-box systems. Aichernig et al.~\cite{aichernig2021learning} used their proposed learning-based fuzzing technique to identify inconsistencies and potential security vulnerabilities in MQTT brokers effectively.

Steffen and Neubauer~\cite{steffen2011simplified} used LearnLib to conduct systematic experimentation with the Online Conference System (OCS) to explore its behavioral potential and understand its emergent behavior. Specifically, they tested the OCS as a black box system, treating it as a virtual user and systematically interacting with it to infer behavioral models resembling Mealy machines. Bainczyk et al.~\cite{bainczyk2017model} have demonstrated a systematic approach to compare diverse implementations of Todo lists through the ALEX tool. Their examination of 27 stable TodoMVC implementations using a two-phase learning-based approach uncovered seven behavioral outliers.

Meinke and Nycander~\cite{meinke2015learning} utilized LBTest~\cite{meinke2013lbtest} to test the robustness of triCalculate, a counter-party credit risk analysis system with a distributed microservice architecture. They injected faults into the system and generated test cases using LBTest to find errors because of those fault injections.  Khan and Sindhu~\cite{khan2022debugging} presented an empirical study evaluating the effectiveness and efficiency of the proposed LBT framework for reactive systems in~\cite{meinke2011incremental} compared to other model-based testing (MBT) tools in debugging software. They specifically evaluated LBT's performance in debugging two case studies of reactive systems: a Cruise Control System (CCS) and an Automated Teller Machine (ATM).  Using two case studies of reactive systems, the study involves 20 participants from various backgrounds in computer science. LBT, employing a black-box learning-based testing framework, outperformed GraphWalker and OSMO Tester in terms of effectiveness and efficiency. The study highlights LBT's potential to improve software development processes and suggests further validation and enhancement. 

Quddus and Sindhu~\cite{quddus2022structural} also evaluated their proposed LBT method using the same two case studies. The authors analyzed the results to determine the extent of structural coverage achieved by the LBT-generated test suite. They found that the test suite provided complete structural coverage of the safety LTL requirements in terms of Requirement Coverage (RC), Antecedent Coverage (AC), and Unique First Cause Coverage (UFCC). However, the structural coverage for liveness LTL requirements was relatively lower, likely due to the complexity introduced by loops in the tests.

Khosrowjerdi et al.~\cite{khosrowjerdi2017automated} applied LBTest to test four case studies provided by industrial partners, Scania CV AB and Volvo Technology Corporation AB: 1) BBW, 2) Remote engine start (ESTA), 3) dual circuit steering (DCS), 4) fuel level display (FLD). They compared LBTest's performance against an industrial testing tool called piTest regarding mutation testing, where LBTest outperformed the established testing framework, piTest, in detecting bugs. Meinke~\cite{meinke2017learning} applied the multi-core version of LBTest, an improved and more scalable iteration, to conduct testing on a vehicle platooning simulator. Zhang et al.~\cite{zhang2017integration}  tested a simple cruise controller with finite-state behavior and a distributive BBW system with continuous data types to conduct an experiment with their proposed LBT method integrated with SCT.

Sophia~\cite{sophia2016learning} assessed the effectiveness of LBTest for testing automotive electronic control units (ECUs). by examining requirement formalization, behavior modeling, and error detection. They found that while 23\% of requirements couldn't be formalized for black box testing, most remaining ones needed reformulation due to issues like missing input assumptions and contradictory scenarios. LBTest proved useful, especially for fast, reactive embedded systems, and demonstrated strong error detection, identifying eight out of ten injected errors compared to the current framework, piTest. However, the tool's effectiveness is contingent on the structure, reactivity, and speed of ECU functions, as well as clear, structured requirements. Improvements in requirement formulation and other areas are necessary for LBTest's effective industry use. khosrowjerdi~\cite{khosrowjerdi2018virtualized} presented a case study for fault injection (FI) testing in automotive embedded systems using a toolchain that integrates QEMU hardware
emulator and GNU debugger GDB with LBTest. The approach aims to evaluate the robustness and safety of ECU software under various fault conditions. Two case studies—Remote Engine Start (ESTA) and Scheduled Memory Corruption Detection (SMCD)—demonstrate the toolchain's capabilities in error discovery, model learning, and performance efficiency. 

Aichernig et al.~\cite{aichernig2019learning} conducted a mutation analysis on AVL489 exhaust measurement device, a non-deterministic system, using LearnLib to check the fault detection capability of the framework. Meinke and Khosrowjerdi~\cite{meinke2021use} evaluated ROBOTest, the CAML approach built on LBTest, by conducting use case testing on an embedded automotive Advanced Driver Assistance System (ADAS) application, demonstrating its efficiency and effectiveness. 

Waga~\cite{waga2020falsification} evaluated their proposed LBT method FalCAuN based on LearnLib by conducting experiments using a Simulink model of an automatic transmission system as the CPS model and various STL formulas as specifications. Test generation involved techniques such as hill climbing, genetic algorithms, and random sampling, while model checking was performed using techniques like the TTT algorithm. The experiments were conducted multiple times to measure the number of falsified specifications and the time taken for falsification. The results were compared across different methods, highlighting the performance of robustness-guided BBC with genetic algorithms in terms of both effectiveness and efficiency. 

LBT techniques, while often applied in specific domains for case study, are versatile and effective for testing a variety of systems across different fields. For example, Raffelt et al. demonstrated LearnLib's~\cite{raffelt2005learnlib} capability to systematically learn finite state machine models of real-world systems. It effectively adapted to the specific requirements and complexities of web applications and telephony hardware, routers, and bug tracking systems~\cite{raffelt2009dynamic} showcasing its versatility and applicability across different domains. AALpy~\cite{muvskardin2022aalpy} has been evaluated through experiments showcasing its performance in learning deterministic and stochastic models, as well as its application in various domains such as fuzzing BLE protocols, model-based diagnosis, extracting models from recurrent neural networks, and debugging software like Vim. In experiments, AALpy demonstrated competitive performance compared to existing libraries like LearnLib. Feng et al.~\cite{feng2013case} employed LBTest to evaluate three commercial software: FAS, an e-commerce access server from Fredhopper; Brake-by-wire system (BBW), an embedded vehicle application featuring Anti-lock Braking Systems (ABS) function from Volvo Technology; and triReduce, a portfolio compression service developed using the popular web framework Django, from TriOptima. 
Notably, LBTest successfully identified bugs in all of these systems, including FAS, which had been developed over a significant period, making it less likely to have bugs. The successful application of LBT techniques in these scenarios validated their effectiveness at integrating model inference and testing.

\section{Discussion}
A notable observation from our analysis of these papers is the gradual evolution of LBT research. 
In its early stages, the theoretical approaches predominantly concentrated on testing procedural programs~\cite{papadopoulos2015black, fraser2015assessing, walkinshaw2017uncertainty,meinke2011learning}. 
However, this emphasis gradually shifted towards testing reactive systems~\cite{meinke2011incremental,meinke2013lbtest,10.1007/978-3-642-24580-0_8}. 
Consequently, nearly all case studies we came across primarily revolved around testing reactive systems using LBT tools~\cite{meinke2011learning,walkinshaw2010increasing,feng2013case}. Recently, LBT methods have also been utilized to test ML models as well models~\cite{muskardin2022learning,sharma2021mlcheck}.

There is a growing interest in cyber-physical systems and networking protocol testing, as the case studies suggest. This could be a future direction of research for LBT, as autonomous vehicle testing has recently been getting more attention than before.

Another discernible transformation pertains to the initial focus on sequential systems, leading to the majority of LBT techniques employing state machines for model inference~\cite{raffelt2005learnlib, shahbaz2009inferring, walkinshaw2009iterative, 10.1007/978-3-642-24580-0_8}. 

However, with the rise in demand for testing data-driven programs, LBT model inference began to incorporate machine learning~\cite{papadopoulos2015black, fraser2015assessing}, genetic algorithms~\cite{fraser2012behaviourally, walkinshaw2017uncertainty}, and other data-driven model inference techniques.
In addition, the LBT method has started being employed in new testing domains such as GUI~\cite{novella2019automatic,choi2013guided}, ML and DNN model testing~\cite{sharma2021mlcheck}.

LBT is not only confined to finding bugs. Due to its incremental learning technique, it has the potential to check robustness in reactive systems in case of fault injection~\cite{meinke2015learning}, which could be a new domain for LBT usability research. 
LBT enhances behavioral adequacy by ensuring that the SUT is comprehensively understood and evaluated. This thorough approach helps uncover hidden defects that traditional testing methods might miss, leading to higher software quality and reliability. By integrating learning and testing, LBT provides a more detailed and accurate assessment of software behavior.

Interestingly, we also observed that LBT can prove effective even without model checking despite being considered an integral component of LBT. 
For instance, Walkinshaw et al.~\cite{walkinshaw2010increasing} demonstrated that LBT can outperform random testing without employing model checkers to generate counterexamples. 

LBT significantly reduces the manual effort required in the testing process by automating many tasks. This automation allows developers and testers to focus on other critical aspects of the development lifecycle, enhancing overall efficiency and effectiveness. The reduced need for manual intervention also minimizes human error, contributing to more reliable testing outcomes.

Unlike traditional model-based software testing, which uses existing design models for generating test cases, LBT infers models directly from the SUT using test data. This approach is beneficial in situations where continuous integration is required to reflect code changes during the implementation process.

\section{Conclusion}
This paper discusses the existing literature, shedding light on the progressive evolution and application of LBT in commercial software testing scenarios. 
Within this context, we explore the various approaches encompassing the three core modules of LBT: 1) model inference, 2) model checking, and 3) test data generation. 
Notably, while the early stages primarily relied on state machines for model inference in theoretical approaches, the contemporary surge in data-driven programs has ushered in the integration of machine learning in this domain. 
Additionally, we observe a discernible shift from the initial focus on testing simple procedural programs to a current emphasis on the testing of intricate embedded reactive systems. 
This shift holds significant implications for the commercial viability of LBT testing. 

The objective of this research is to provide fresh insights for researchers in the field of LBT and to pave the way for new horizons in software testing at large. We highlighted successful implementations and best practices through systematic reviews that encourage the adoption of effective LBT techniques, leading to more consistent and reliable testing processes across the software industry by ensuring that proven methods are widely recognized and utilized. This paper also provides a ground for future research by consolidating existing knowledge and raising awareness about the benefits and potential of LBT, further promoting its adoption, streamlining research efforts, and ensuring that new research builds on a solid foundation of prior achievements.

\bibliographystyle{plainnat} 
\bibliography{sigproc}








 



\end{document}